\documentclass[aps,pra,twocolumn,superscriptaddress,showpacs]{revtex4}
\usepackage{amsmath,amssymb,graphicx}
\usepackage[hypertex,breaklinks=true,colorlinks=false]{hyperref}

\newcommand{\PRL}[3]{Phys. Rev. Lett. {\bf #1},
\href{http://link.aps.org/abstract/PRL/v#1/e#2}{#2} (#3)}
\newcommand{\PRLp}[3]{Phys. Rev. Lett. {\bf #1},
\href{http://link.aps.org/abstract/PRL/v#1/p#2}{#2} (#3)}
\newcommand{\PRB}[3]{Phys. Rev. B {\bf #1},
\href{http://link.aps.org/abstract/PRB/v#1/e#2}{#2} (#3)}
\newcommand{\PRBp}[3]{Phys. Rev. B {\bf #1},
\href{http://link.aps.org/abstract/PRB/v#1/p#2}{#2} (#3)}
\newcommand{\PRBR}[3]{Phys. Rev. B {\bf #1},
\href{http://link.aps.org/abstract/PRB/v#1/e#2}{#2(R)} (#3)}
\newcommand{\PRX}[3]{Phys. Rev. X {\bf #1},
\href{http://link.aps.org/abstract/PRX/v#1/e#2}{#2} (#3)}
\newcommand{\RMP}[3]{Rev. Mod. Phys. {\bf #1},
\href{http://link.aps.org/abstract/RMP/v#1/p#2}{#2} (#3)}

\begin{document}
\title{Phase diagram and pair Tomonaga-Luttinger liquid\\ 
in a Bose-Hubbard model with flat bands}
\author{Shintaro Takayoshi}
\affiliation{National Institute for Materials Science, Tsukuba 305-0047, Japan}
\author{Hosho Katsura}
\affiliation{Department of Physics, Gakushuin University, Tokyo 171-8588, Japan}
\author{Noriaki Watanabe}
\affiliation{Department of Physics, The University of Tokyo, Tokyo 113-0033, Japan}
\affiliation{Kavli Institute for the Physics and Mathematics of the Universe,
The University of Tokyo, Kashiwa, Chiba 277-8583, Japan}
\author{Hideo Aoki}
\affiliation{Department of Physics, The University of Tokyo, Tokyo 113-0033, Japan}

\date{\today}

\begin{abstract}
To explore superfluidity in flat-band systems, 
we consider a Bose-Hubbard model on a cross-linked ladder with $\pi$ flux, 
which has a flat band with a gap between the other band 
for noninteracting particles, 
where we study the effect of the on-site repulsion nonperturbatively. 
For low densities, we find exact degenerate ground states, 
each of which is a Wigner solid with nonoverlapping 
Wannier states on the flat band. 
At higher densities, the many-body system, when projected onto the lower 
flat band, can be mapped to a spin-chain model. 
This mapping enables us to reveal the existence of a 
Tomonaga-Luttinger liquid comprising {\it pairs of bosons}. 
Interestingly, the high- and low- density regions have an overlap, 
where the two phases coexist. 
\end{abstract}


\maketitle

\section{Introduction}
The macroscopically degenerate manifolds of single-particle states 
on a flat band provide a unique playground, especially 
for studying non-perturbative aspects of electron correlations 
in fermionic lattice models. 
For example, Lieb, and subsequently Mielke and Tasaki, have shown rigorously that 
repulsive Hubbard interaction provokes 
ferri- or ferro-magnetism~\cite{Lieb89,Mielke91,Tasaki92}, 
and localized magnons are discussed 
in frustrated antiferromagnets~\cite{Schulenburg02,Zhitomirsky04}. 
In a much wider context, flat-band systems have attracted a renewed 
attention in terms of topological insulators~\cite{Hasan10,Qi11}, 
where several constructions were proposed for exactly or nearly 
flat bands that accommodate nontrivial Chern numbers~\cite{Katsura10,Tang11,Sun11,Neupert11}. 
These topological flat bands are reminiscent of 
Landau levels in the conventional systems in magnetic field 
and may set the stage for new topological phases when interactions are switched on. 
Indeed, the Hubbard interaction has been shown to lead to 
either conventional ferromagnetic states~\cite{Katsura10,Neupert12} or 
fractional Chern insulators~\cite{Sheng11,Regnault11}, 
depending on the fermion filling of the flat band. 

\begin{figure}[t]
\includegraphics[width=0.4\textwidth]{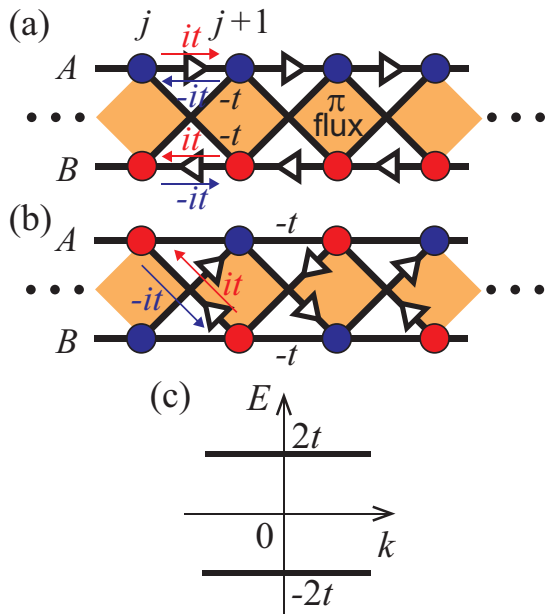}
\caption{(a) Bose-Hubbard model on a cross-linked ladder with $\pi$ flux. 
Circles represent sites. $A$ (blue), $B$ (red), and 
$\ldots,j-1,j,j+1,\ldots$ are chain 
and rung indices, respectively. While interchain diagonal hopping is $-t$, 
intrachain hopping is $it$ ($-it$) 
for along (against) the direction of the arrow, respectively. 
(b) The same model with interchange of A and B sites on rungs with $j={\rm even}$. 
(c) Dispersion relation of the present model for the noninteracting case.
It has two flat bands at $E=\pm 2t$.}
\label{fig:model}
\end{figure}

In the present work, we consider a {\it bosonic} system with a Hubbard interaction 
on a flat-band lattice, which is a cross-linked ladder 
with $\pi$ flux [Figs.~\ref{fig:model}(a) and \ref{fig:model}(b)].  
The basic motivation for going over to a bosonic system is the following. 
In the fermionic flat-band systems, 
one important observation is that correlated systems 
in a flat band are quite distinct from 
those in the atomic limit (where the hopping in the 
tight-binding model $\to 0$)~\cite{Kusakabe94}. 
Then, an intriguing question arises: 
Can there be Bose-Einstein condensation (BEC) in flat-band boson systems 
while we have obviously no BEC
in the atomic limit (with an effective mass $\to\infty$)? 
If we do have states far from those in the atomic limit, 
we may have a peculiar quantum liquid, 
or even a possibility of supersolid, 
i.e., coexistence of condensation with a diagonal long-range order 
such as charge density wave (CDW).  
In fact, Huber and Altman have considered this problem, 
and they concluded that there is a supersolid (superfluid + CDW) region 
for a chain of triangle and kagome lattices~\cite{Huber10}. 

Our choice of the model [Fig.~\ref{fig:model}(a)] 
is inspired by the work of Creutz {\it et al}.~\cite{Creutz99,Bermudez09}, 
which shows that edge states emerge at open boundaries
and the fermionic counterpart can be thought of 
as one-dimensional topological insulators~\cite{Mazza12}. 
In this paper, we find that the many-body ground state (GS) exhibits 
a variety of phases as the density of bosons is varied. 
While the GS consists of nonoverlapping Wannier states at low densities, 
we find that Tomonaga-Luttinger liquid (TLL) 
comprising pairs of bosons (pair TLL) is stabilized at higher densities, 
where the fundamental degrees of freedom are boson pairs 
rather than individual bosons. 
Pair TLL is deduced from the mapping of a boson system 
onto a spin-1/2 XXZ chain in magnetic field. 
The validity of the mapping is confirmed by exact diagonalization 
and the Bethe ansatz solution. 
The analysis of the spin chain also reveals the coexistence of 
Wigner solid and pair TLL phases at intermediate densities. 

\section{Formulation}
We consider a cross-linked ladder [see Fig.~\ref{fig:model}(a)] 
with a Hamiltonian,
\begin{align}
{\cal H} &= {\cal H}^{\rm kin} + 
\frac{U}{2}\sum_{j=1}^{L}\sum_{X=A,B}b_{j}^{X\dagger}b_{j}^{X\dagger}b_{j}^{X}b_{j}^{X},
\nonumber \\
{\cal H}^{\rm kin} &= -t\sum_{j=1}^{L}(e^{ i\theta}b_{j}^{A\dagger}b_{j+1}^{A}
                           +e^{-i\theta}b_{j}^{B\dagger}b_{j+1}^{B}\nonumber\\
              &\qquad\qquad+            b_{j}^{A\dagger}b_{j+1}^{B}
                           +            b_{j}^{B\dagger}b_{j+1}^{A}+{\rm H.c.}),
\label{eq:Hamil_orig}
\end{align}
with an appropriate local gauge, 
where $b_{j}^{X\dagger}$ creates a boson, 
$j$ and $X$ ($=A$, $B$) represent indices 
for rungs and legs of the ladder, respectively, and 
$L$ is the length of the ladder (that is, the number of rungs), 
so the total number of sites is $2L$. 
The boundary condition is periodic.
It is the same as the Creutz model~\cite{Creutz99} 
except that particles are not fermions but bosons. 
Here we focus on the case of $\theta=\pi/2$, 
which means $\pi$ flux is introduced. 
Figure~\ref{fig:model}(b) is the same model 
with interchanging A and B sites on rungs with $j={\rm even}$. 
It is useful for analyzing (\ref{eq:Hamil_orig}) to take the Wannier basis 
\begin{equation}
 w_{j}^{\pm}=\frac{1}{2}[ib_{j}^{A}- b_{j}^{B}
                    \pm(b_{j+1}^{A}-ib_{j+1}^{B})],
\end{equation}
which satisfies bosonic commutation relations: 
$[w_{j}^{\alpha},w_{j'}^{\alpha' \dagger}]=\delta_{\alpha,\alpha'}\delta_{j,j'}$, 
$[w_{j}^{\alpha},w_{j'}^{\alpha'}]=[w_{j}^{\alpha\dagger},w_{j'}^{\alpha' \dagger}]=0$. 
The kinetic part of the Hamiltonian~(\ref{eq:Hamil_orig}) becomes 
${\cal H}^{\rm kin}=\sum_{j=1}^{L}\sum_{\alpha=\pm}
\epsilon_{\alpha}w_{j}^{\alpha\dagger}w_{j}^{\alpha}$, 
where $\epsilon_{\pm}=\pm 2t$. 
The two energy bands are both flat bands 
having no wave number dependence (no dispersion), 
as shown in Fig.~\ref{fig:model}(c).

\section{Results}
\begin{figure}[t]
\includegraphics[width=0.48\textwidth]{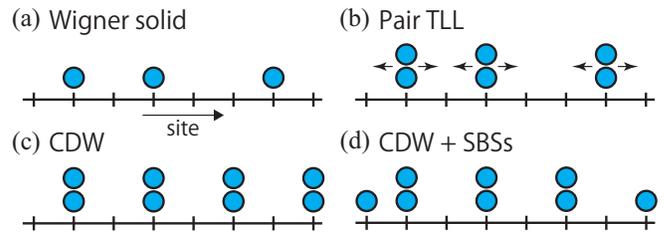}
\caption{Schematic pictures of (a) Wigner solid, 
(b) pair TLL, (c) CDW, and 
(d) CDW with two single-boson sites (SBSs). 
Particles are localized in panels (a), (c), and (d) 
while they are mobile (arrows) in panel (b).}
\label{fig:phases}
\end{figure}

Since the kinetic energy is quenched in the above sense, 
we have only to consider the particle-particle interaction.  
The Wannier operator $w_{j}^{\alpha}$ involves $j$ and $j+1$ sites, 
so that they have only on-site and 
nearest-neighbor interactions. 
When the boson density (the averaged number of bosons per site) 
$\rho\equiv N/(2L)$ is dilute ($\rho \leq 1/4$), 
we can construct a state that 
does not feel the interaction by making configurations in which 
the Wannier state at each site (which we simply call ``site'' hereafter) 
is occupied by 0 or 1 particle 
with no two nearest-neighbor sites occupied [Fig.~\ref{fig:phases}(a)]. 
Since the kinetic energy is quenched and the interaction term 
is positive semidefinite, these states are obviously many-body 
GS of the Hamiltonian ${\cal H}$.
This configuration resembles the Wigner solid in an electron gas 
that minimizes the Coulomb interaction by avoiding overlap, 
so we call it a Wigner solid hereafter.

For higher density region of $\rho>1/4$, 
we consider a projection of Hamiltonian onto 
the lower energy band $\alpha=-$, neglecting the effect of 
the higher energy band $\alpha=+$ 
(interband interaction and intraband interaction within the higher band). 
This approximation is justified when the interaction $U$ 
is small compared with the hopping $t$. 
We then obtain 
\begin{align}
 {\cal H}_{\rm proj}^{\rm int}
    & = \sum_{j=1}^{L} \left[\frac{U}{8} w_{j}^{\dagger}w_{j}^{\dagger}w_{j}w_{j} 
    +\frac{U}{4} w_{j}^{\dagger}w_{j+1}^{\dagger}w_{j}w_{j+1}\right. \nonumber\\
    &-\frac{U}{16}(w_{j}^{\dagger}w_{j}^{\dagger}w_{j+1}w_{j+1}
                     +\left.{\rm H.c.})\right],
\label{eq:Hamil_proj}
\end{align}
where we have dropped the band index ``$-$''. 
Namely, we end up with 
on-site repulsion (first term on the right-hand side), 
nearest-neighbor repulsion (second), and 
pair hopping of bosons (third), respectively.  
Remarkably, we have a pair-hopping interaction with a 
significant magnitude ($U/16$). 
The sign of this term is irrelevant, 
since it is altered by a gauge transformation, 
$w_{j}\to iw_{j}$ (for $j={\rm even}$). 
We can vary the number of bosons $N$ by adding the chemical potential term 
$-\mu N\equiv -\mu\sum_{j=1}^{L}w_{j}^{\dagger}w_{j}$ to~(\ref{eq:Hamil_proj}). 
In the following, we ignore the states 
in which more than two bosons occupy a single site 
because the energy would become much higher due to the on-site repulsion. 
Namely, the number of bosons in a single site is 0, 1, or 2, 
and these states are denoted by $|0\rangle,|1\rangle,|2\rangle$, respectively. 
Note that a single boson is immobile because bosons can hop only in pairs, 
and let us call a site occupied by a single boson a single-boson site (SBS). 
If we only consider $|0\rangle$ and $|2\rangle$, 
Eq.~(\ref{eq:Hamil_proj}) can be mapped to a spin-1/2 model 
by identifying $|0\rangle\mapsto|\downarrow\rangle$ and 
$|2\rangle\mapsto|\uparrow\rangle$. 
We introduce spin-1/2 operators $S_{j}^{x(y,z)}$ 
and $S_{j}^{\pm}=S_{j}^{x}\pm iS_{j}^{y}$, with which we have 
$w_{j}^{\dagger}w_{j}^{\dagger}w_{j}w_{j}\mapsto 2S_{j}^{z}+1$, 
$w_{j}^{\dagger}w_{j+1}^{\dagger}w_{j}w_{j+1}\mapsto (2S_{j}^{z}+1)(2S_{j+1}^{z}+1)$, 
$w_{j}^{\dagger}w_{j}^{\dagger}w_{j+1}w_{j+1}\mapsto 2S_{j}^{+}S_{j+1}^{-}$. 
The Hamiltonian is then mapped to
\begin{align}
 {\cal H}_{S=1/2}
  =&\frac{U}{4}\sum_{j=1}^{L}(-S_{j}^{x}S_{j+1}^{x}-S_{j}^{y}S_{j+1}^{y}+4S_{j}^{z}S_{j+1}^{z}) 
   \nonumber\\
   &\;\;+\frac{5}{4}U\sum_{j=1}^{L} S_{j}^{z}+\frac{3}{8}UL.
\label{eq:Hamil_spinhalf}
\end{align}
This is nothing but the spin-1/2 XXZ model in a magnetic field. 
The chemical potential term is translated to 
$-2\mu\sum_{j=1}^{L}(S_{j}^{z}+1/2)$. 

If we want to include $|1\rangle$, we only have to 
map the system to a spin-1 chain 
($|0\rangle\mapsto|-1\rangle$, 
 $|1\rangle\mapsto| 0\rangle$, 
 $|2\rangle\mapsto| 1\rangle$), with which we have 
$w_{j}^{\dagger}w_{j}^{\dagger}w_{j}w_{j}\mapsto T_{j}^{z}(T_{j}^{z}+1)$, 
$w_{j}^{\dagger}w_{j+1}^{\dagger}w_{j}w_{j+1}\mapsto (T_{j}^{z}+1)(T_{j+1}^{z}+1)$, 
$w_{j}^{\dagger}w_{j}^{\dagger}w_{j+1}w_{j+1}\mapsto \frac{1}{2}(T_{j}^{+})^{2}(T_{j+1}^{-})^{2}$, 
where $T_{j}^{x(y,z)}$ are spin-1 operators 
and $T_{j}^{\pm}=T_{j}^{x}\pm iT_{j}^{y}$.
In this case we have a Hamiltonian 
\begin{align}
 {\cal H}_{S=1}
  =\frac{U}{32}\sum_{j=1}^{L} \left[-(T_{j}^{+})^{2}(T_{j+1}^{-})^{2}
                                -(T_{j}^{-})^{2}(T_{j+1}^{+})^{2}\right.
                                   \nonumber\\
   +\left.8T_{j}^{z}T_{j+1}^{z}\right] + \frac{5}{8}U\sum_{j=1}^{L} T_{j}^{z}
        +\frac{U}{8} \sum_{j=1}^{L}(T_{j}^{z})^{2}
        +\frac{1}{4}UL.
\label{eq:Hamil_spinone}
\end{align}
These mappings help us investigating GS and low energy excitation structure 
of~(\ref{eq:Hamil_proj}) below. 

Now let us begin with the region of $1/4<\rho<1/2$, 
where SBSs are adjacent or two bosons occupy the same site. 
Since the nearest-neighbor repulsion is stronger than on-site repulsion 
in (\ref{eq:Hamil_proj}), particles tend to form pairs 
rather than to occupy adjacent sites. 
We have performed an exact diagonalization, which demonstrates that 
the GS of (\ref{eq:Hamil_spinhalf}) and (\ref{eq:Hamil_spinone}) numerically 
coincide with each other (see Appendix~\ref{app:EDresult}). 
In other words, SBSs are irrelevant, and the use of ${\cal H}_{S=1/2}$ in place 
of ${\cal H}_{S=1}$ is justified. 
The $1/4<\rho<1/2$ regime for bosons corresponds to 
the spin-1/2 XXZ model~(\ref{eq:Hamil_spinhalf}) 
with nonzero magnetization, 
where the low-energy excitations are known to be described 
as TLL induced by external magnetic fields. 
Since $|\downarrow\rangle$ and $|\uparrow\rangle$ correspond to 
$|0\rangle$ and $|2\rangle$, respectively, we can translate the state 
as a ``pair TLL'' back in the original model. 
A schematic picture of pair TLL is shown in Fig.~\ref{fig:phases}(b).

We can further make a quantitative analysis 
since the spin-1/2 XXZ chain is exactly solvable. 
By solving Bethe ansatz integral equations~\cite{Qin97,Cabra98}, 
we can determine the spinon velocity $v$ and the Luttinger parameter $K$, 
which characterize correlation functions as
\begin{align}
 &\langle n_{r}n_{0}\rangle
   \simeq 4\rho^{2}-2K(\pi r)^{-2}
         +c_{1}\cos(2\pi\rho r)r^{-2K},\nonumber\\
 &\langle w_{r}^{\dagger}w_{r}^{\dagger}w_{0}w_{0}\rangle
   \simeq (-)^{r}c_{2}r^{-\frac{1}{2K}}
         +(-)^{r}c_{3}\cos(2\pi\rho r)r^{-2K-\frac{1}{2K}}.\nonumber
\end{align}
Here the lattice constant is taken to be unity, 
$n_{r}=w_{r}^{\dagger}w_{r}$, and 
$c_{1},c_{2},c_{3}$ are nonuniversal constants. 
In the present case, where $1/4\leq K\leq 1/2$, 
$\langle w_{r}^{\dagger}w_{r}^{\dagger}w_{0}w_{0}\rangle$ 
decays faster than $\langle n_{r}n_{0}\rangle$.
Note that collective excitations in TLL are gapless 
in contrast to the gapped excitations for creating SBSs. 
This fact again justifies the neglect of SBSs in 
taking the low-energy effective Hamiltonian. 

\begin{figure}[t]
\includegraphics[width=0.48\textwidth]{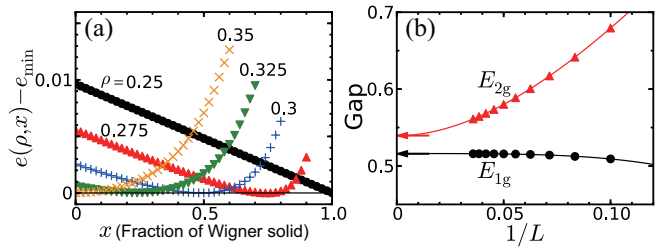}
\caption{(a) The GS energy per site above its minimum, 
$e(\rho,x)-e_{\rm min}$, against the fraction of Wigner solid phase $x$ 
in the density region $0.25\leq\rho\leq 0.35$. 
(b) System-size ($L$) dependence of 
the energy of the state with two adjacent SBSs ($E_{1{\rm g}}$) and 
that of the first excited state without SBSs ($E_{2{\rm g}}$) 
at $\rho=1/2$. 
Solid curves represent the extrapolation 
by fourth-order polynomials, while arrows are 
the analytically obtained values. 
The energy unit is $U=1$.}
\label{fig:energylevel}
\end{figure}

Here we can raise an intriguing question: 
Can Wigner solid and pair TLL phases coexist? 
Since each boundary between the two phases costs 
surface energy, there should be a single boundary, if any. 
We assume that the proportion of Wigner solid phase is $x \;\;(0\leq x\leq 1)$, 
i.e., Wigner solid phase has $xL/2$ bosons occupying $2xL$ sites, 
while pair TLL phase has $N- xL/2$ bosons occupying $2(1-x)L$ sites. 
This pair TLL translates into the spin-1/2 chain with magnetization 
$M=1/2-(\rho-x/4)/(1-x)$ 
(with the saturated magnetization corresponding to 1/2), 
and the GS energy per site $e_{\rm TLL}(\rho,x)$ can be obtained 
by solving the Bethe-ansatz equations~\cite{Qin97,Cabra98}. 
The energy in the total system is 
$e(\rho,x)\equiv(1-x)e_{\rm TLL}(\rho,x)$, 
since the Wigner solid phase has zero energy. 
We can then determine $x$ by minimizing $e(\rho,x)$. 
Figure ~\ref{fig:energylevel}(a) displays 
$e(\rho,x)-e_{\rm min}$ ($e_{\rm min}$: the minimum value) against $x$. 
We find that 
for $\rho\leq 0.25$ the minimum occurs at $x=1$ 
(a pure Wigner solid phase), while for $\rho\geq 0.35$ 
the minimum locates at $x=0$ (a pure pair TLL phase). 
In the intermediate regions ($0.25\leq\rho\leq 0.35$) 
we do have a coexistence of the two phases (with $0<x<1$). 

Finally, we consider the $\rho=1/2$ case. 
Since the GSs of~(\ref{eq:Hamil_spinhalf}) and~(\ref{eq:Hamil_spinone}) 
coincide according to numerical diagonalization,  
the spin-1/2 description is still valid. 
However, we have to be careful about the excited states involving SBSs 
because the $\rho=1/2$ case corresponds to the XXZ chain 
without magnetization and has an excitation gap. 
The lowest two levels are doubly degenerate GSs, 
which have a N\'eel order due to a strong Ising anisotropy 
in the XXZ description~(\ref{eq:Hamil_spinhalf}). 
This state translates to a charge density wave (CDW) of 
pair bosons [Fig.~\ref{fig:phases}(c)].
Let us consider the configuration that 
contains two adjacent SBSs [Fig.~\ref{fig:phases}(d)]. 
It translates into a system of length $L-2$ having 
$N-2$ bosons with open boundary condition, 
for which the Hamiltonian is Eq.~(\ref{eq:Hamil_spinhalf}) 
with the summation changed to $\sum_{j=2}^{L-1}$ 
(with $j=1$ and $L$ occupied by SBSs) and 
a constant ($U/4$) subtracted. 
We have confirmed that the two lowest energy levels of this open chain 
agree with the lowest two excitations. 
These levels correspond to CDW of $N-2$ bosons (N\'eel state 
in the spin model) in a chain of length $L-2$. 

Upon a closer look, 
the energy of the state with two adjacent SBSs ($E_{1{\rm g}}$) and 
that of the first excited state without SBSs ($E_{2{\rm g}}$) 
are calculated by exact diagonalization for various system sizes, 
which is shown in Fig.~\ref{fig:energylevel}(b). 
The thermodynamic limit of $E_{1{\rm g}}$ and $E_{2{\rm g}}$ 
is determined exactly with the Bethe ansatz~\cite{Batchelor90,Kapustin96} as
$E_{1{\rm g}}^{(L\to\infty)}\simeq 0.515877$ and 
$E_{2{\rm g}}^{(L\to\infty)}\simeq 0.539548$ (see Appendix~\ref{app:BetheAnsatz}). 
Here, we take the energy unit as $U=1$. 
If we extrapolate $E_{1{\rm g}}$ and $E_{2{\rm g}}$ 
with fourth-order polynomials [solid curves in Fig.~\ref{fig:energylevel}(b)], 
the thermodynamic limit agrees with the Bethe ansatz result accurately. 
Thus, it is confirmed that the low-energy structure, 
except for the midgap state (corresponding to $E_{1{\rm g}}$), 
is described well by the effective spin-1/2 model~(\ref{eq:Hamil_spinhalf}). 

\begin{figure}[t]
\includegraphics[width=0.48\textwidth]{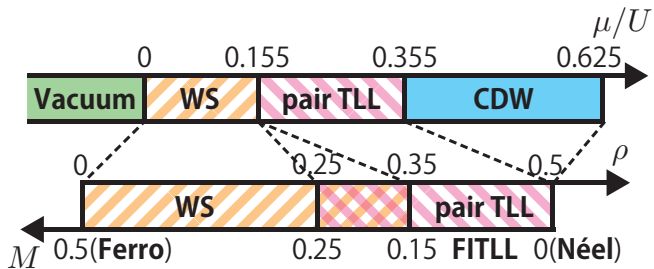}
\caption{GS phase diagram as a function of 
the chemical potential $\mu$ or the density $\rho$. 
WS and FITLL stand for Wigner solid and field-induced TLL, respectively. 
The bottom axis is the corresponding phase diagram 
for the spin-1/2 XXZ chain as a function of magnetization $M$.}
\label{fig:phasediag}
\end{figure}

Now we can summarize the phase diagram in Fig.~\ref{fig:phasediag}. 
The transition from Wigner solid to the pair TLL is of the first order, 
accompanied by a region with coexistence of the two phases. 
The transition from pair TLL to CDW translates to that 
from field-induced TLL phase to N\'eel state in spin-1/2 XXZ model, 
which belongs to 
Dzhaparidze-Nersesyan-Pokrovsky-Talapov universality class~\cite{Cabra98}. 
The relation between chemical potential $\mu$ and boson density $\rho$ 
in the pair TLL phase corresponds to the $M$-$H$ curve 
(magnetization vs magnetic field) in the XXZ model. 

\section{Summary and discussion}
We have studied how the repulsive interaction exerts 
an effect on bosons in a flat-band Bose-Hubbard model 
on a cross-linked ladder with $\pi$ flux. 
In the low boson-density region $\rho\leq 1/4$, 
the GS is Wigner solid with nonoverlapping Wannier states. 
We have revealed an existence of pair TLL when $\mu$ is increased, 
after the first-order transition accompanying a phase coexistence region. 
Finally, the GS becomes CDW at $\rho=1/2$. 
The phase diagram and low-energy excitation structure are analyzed 
through the mapping to an effective spin-1/2 XXZ chain, 
which is confirmed by exact diagonalization and the Bethe ansatz. 

Our model is relevant to a Bose-Hubbard model on a diamond chain, 
which has been realized experimentally~\cite{Gladchenko09}. 
A pair TLL is also pointed out in this diamond chain 
with $\pi$ flux~\cite{Vidal00,Doucot02} as a ``nematic Luttinger liquid," 
but they employed the mapping to the classical rotator model. 
Their model has three flat bands~\cite{Gulacsi07} at $E=0,\pm 2t$, 
and its analysis can be done in the same way as we do here. 
If the Hamiltonian is projected onto the lowest band with $E=-2t$, 
the on-site repulsion is stronger than the nearest-neighbor one. 
Thus the system is always in the Wigner solid phase while the pair TLL does not appear. 
The projection to the $E=0$ band would provide a model similar to 
the effective model for the cross-linked ladder, 
but such a projection would not be justified, 
since the effect of $E=\pm 2t$ bands is 
not negligible in both low- and high-boson-density regions. 
We can prove that the lower band in our model corresponds to 
$E=0$ band in the diamond chain by integrating out the vertex having 
four edges of the latter model. 
Therefore, we believe that our approach is more rigorous 
in demonstrating the existence of the pair TLL phase. 

Recent progress in cold atom physics has enabled experimentalists to 
realize bosons with ``synthetic gauge fields"~\cite{Lin09}, 
and our result is expected to be tested with bosons trapped in 
an optical lattice with a ladder configuration. 
While we have concentrated here on a bosonic system, 
searching for nontrivial phases 
in fermionic flat-band systems~\cite{Filho08,Lopes11} 
is an interesting future problem. 
If two chains are considered as spin up and spin down, 
Hamiltonian (\ref{eq:Hamil_orig}) may be regarded as a one-dimensional 
electron system with a spin-orbit coupling. 
Also, solving a problem by mapping to spin chains 
reminds us of the Tao-Thouless (thin torus) limit~\cite{Tao83,Nakamura10}
of the fractional quantum Hall system. 
Considering higher-dimensional flat-band systems is 
another direction to study~\cite{Huber10}. 

\begin{acknowledgements}
H.A. is supported by a Grant-in-Aid from JSPS No. 23340112, 
and H.K. is supported by a Grant-in-Aid for Young Scientists (B) No. 23740298.
\end{acknowledgements}

\appendix
\section{Justifications of the mapping \\to the spin-1/2 chain} 
\label{app:EDresult}
In order to study the Hamiltonian~(\ref{eq:Hamil_proj}),
we have exploited the mapping to a spin chain model. 
When SBSs are neglected, 
the mapped model is a spin-1/2 chain 
with a Hamiltonian~(\ref{eq:Hamil_spinhalf})
while if we include SBSs we have an $S=1$ model~(\ref{eq:Hamil_spinone})

\begin{table}[t]
\begin{center}
\caption{The GS energies for the 
spin models~(\ref{eq:Hamil_spinhalf}) and~(\ref{eq:Hamil_spinone}), 
respectively, of length $L=16$ calculated with exact diagonalization. 
The GS energy is shown in the unit of $U=1$.}
\label{tab:GSenergy}
\begin{tabular}{|c|c||c|c|}
\hline
 $N$ &   $\rho$   & GS energy of~(\ref{eq:Hamil_spinhalf}) & GS energy of~(\ref{eq:Hamil_spinone}) \\\hline
   8 & \;0.2500\; & \;0.1453388529\; & \;0.0000000000\; \\
  10 &   0.3125   &   0.3214844344   &   0.3214844344   \\
  12 &   0.3750   &   0.6206828185   &   0.6206828185   \\
  14 &   0.4375   &   1.0891052450   &   1.0891052450   \\
  16 &   0.5000   &   1.7536447794   &   1.7536447794   \\\hline
\end{tabular}
\end{center}
\end{table}

To justify the neglect of SBSs, 
we have performed an exact diagonalization. 
The GS energies of 
(\ref{eq:Hamil_spinhalf}) and (\ref{eq:Hamil_spinone})
of length $L=16$ are respectively 
shown in Table~\ref{tab:GSenergy}. 
For the number of bosons $N > L/2$ ($\rho>1/4$), Hamiltonians 
(\ref{eq:Hamil_spinhalf}) and (\ref{eq:Hamil_spinone}) 
are seen to indeed share the same GSs. For 
$0 \le N \le L/2$ ($0 \le \rho \le 1/4$), on the other hand, 
the GS of~(\ref{eq:Hamil_spinone}) is 
the zero-energy state (Wigner solid), 
which is not the GS of~(\ref{eq:Hamil_spinhalf}).
In Table~\ref{tab:GSenergy}, $N=10$ $(\rho=5/16)$ should correspond to the coexistence region, 
but the energy of the coexistence state is raised due to 
the surface energy in finite systems, 
and the pair Tomonaga-Luttinger liquid (pair TLL) becomes the GS. 

\section{Low-energy excitation structure \\at $\rho=1/2$}

\begin{table}[b]
\begin{center}
\caption{Several lowest energy levels of
$S=1$ model~(\ref{eq:Hamil_spinone})
with $L=16$ calculated with exact diagonalization. 
The GS energy is shown in the unit of $U=1$.}
\label{tab:S1excitation}
\begin{tabular}{|c||c|c|c|}
\hline
      &    Energy        & Degeneracy & Number of SBSs  \\\hline
\;1\; & \;1.7536447794\; &      1     &      0          \\
  2   &   1.7541665514   &      1     &      0          \\
  3   &   2.2686557182   &     16     &      2          \\
  4   &   2.2709315403   &     16     &      2          \\
  5   &   2.3541625572   &      1     &      0          \\
  6   &   2.3668727422   &      1     &      0          \\
  7   &   2.4029012587   &      2     &      0          \\
  8   &   2.4115680946   &      2     &      0          \\\hline
\end{tabular}
\end{center}
\end{table}

For the case of $\rho=1/2$, the spin-1/2 description is still valid, 
since the GSs of~(\ref{eq:Hamil_spinhalf}) and~(\ref{eq:Hamil_spinone}) 
coincide with each other for $N=16$ as seen in Table~\ref{tab:GSenergy}. 
However, we have to pay attention to the excited states involving SBSs 
because the $\rho=1/2$ case corresponds to the XXZ chain 
without magnetization for which an excitation gap exists. 
Several lower-most levels of the spin-1 model~(\ref{eq:Hamil_spinone}) 
are displayed in Table~\ref{tab:S1excitation}. 
The lowest two levels are doubly-degenerate GSs 
(although 
the degeneracy is slightly lifted due to a finite-size effect), 
which have a N\'eel order due to the strong Ising anisotropy 
in the XXZ description~(\ref{eq:Hamil_spinhalf}). 
This state translates to a charge density wave (CDW) of pair bosons.

The first two excited levels in Table~\ref{tab:S1excitation} 
do not appear in the energy levels of~(\ref{eq:Hamil_spinhalf}), 
and hence they should correspond to the configuration involving SBSs. 
The system having two adjacent SBSs at site $j=1$ and $L$ 
can be mapped to a 
spin-1/2 chain with open boundary conditions such as 
\begin{align}
 {\cal H}_{S=1/2}^{\rm open}
  =& \frac{U}{4}\sum_{j=2}^{L-2}(-S_{j}^{x}S_{j+1}^{x}-S_{j}^{y}S_{j+1}^{y}
                                +4S_{j}^{z}S_{j+1}^{z})\nonumber\\ 
   &+\frac{5}{4}U\sum_{j=2}^{L-1} S_{j}^{z}+\frac{3}{8}UL-\frac{U}{4}.
\label{eq:Hamil_spinhalf_open}
\end{align}
We have confirmed that the two lowest energy levels of~(\ref{eq:Hamil_spinhalf_open}) do 
agree with the first two excited states in Table~\ref{tab:S1excitation}. 
These levels correspond to CDW of $N-2$ bosons (N\'eel state 
in the spin model) in a chain of length $L-2$ with open boundary condition. 
Each of them is $L$-fold degenerated, since the neighboring 
SBSs have a translational invariance.

\section{Exact results from Bethe ansatz}
\label{app:BetheAnsatz}
As for the energy of the state with two adjacent SBSs ($E_{1{\rm g}}$) and 
that of the first excited state without SBSs ($E_{2{\rm g}}$), 
we can notice that the thermodynamic limit can be calculated exactly 
with the Bethe ansatz~\cite{Batchelor90,Kapustin96}. 
Let us define $\theta \equiv \cosh^{-1}\Delta >0$, 
where $\Delta$ represents the Ising anisotropy of the XXZ chain. 
In the present case $\Delta=4$, hence $\theta=\ln(4+\sqrt{17})$, 
and we have
\begin{align}
E_{1{\rm g}}^{(L\to\infty)}&=-2e_{0}+E_{\rm surf}-1/4\simeq 0.515877,\nonumber\\
E_{2{\rm g}}^{(L\to\infty)}&= \frac{\sinh\theta}{4}\prod_{n=1}^{\infty}
 \left(\frac{{\rm e}^{n\theta}-1}{{\rm e}^{n\theta}+1}\right)^{2}
 \simeq 0.539548,\nonumber
\end{align}
where
\begin{align}
 e_{0}=&\frac{1}{16}\cosh\theta-\frac{1}{8}\sinh\theta
 \left(1+4\sum_{n=1}^{\infty}\frac{1}{{\rm e}^{2n\theta}+1}\right),\nonumber\\
E_{\rm surf}=&-\frac{\Delta}{16}+\frac{1}{2}\sinh\theta
  \left[\frac{1}{4}
       +\sum_{n=1}^{\infty}\frac{{\rm e}^{2n\theta}-1}{{\rm e}^{4n\theta}+1}
       +\sum_{n=1}^{\infty}\frac{(-1)^{n}}{{\rm e}^{2n\theta}+1}\right],\nonumber
\end{align}
represent the ground-state energy per site~\cite{Batchelor90} 
and the surface energy~\cite{Kapustin96} of the XXZ chain, respectively. 
Here, the energy unit $U=1$ is taken.

\end{document}